\documentclass[12pt]{iopart}

\usepackage[dvips]{graphicx}
\usepackage{amssymb}
\usepackage{textcomp}
\usepackage{units}
\usepackage{url}
\usepackage[latin1]{inputenc} 

\newcommand{\CFS}{Co$_2$FeSi(100)}
\newcommand{\ep}{\varepsilon}
\newcommand{\FM}{\mathrm{FM}}
\newcommand{\sat}{\mathrm{sat}}
\newcommand{\subcfs}{\mathrm{CFS}}
\def\vek#1{\mathbf{#1}}

\begin{document}

\title
{Huge quadratic magneto-optical Kerr effect and magnetization
reversal in the Co$_2$FeSi Heusler compound}

\author{J. Hamrle, S. Blomeier, O. Gaier, B. Hillebrands}
\address{Fachbereich Physik and Forschungsschwerpunkt MINAS,
Technische Universit\"at Kaiserslautern,
Erwin-Schr\"odinger-Stra\ss e 56, D-67663 Kaiserslautern, Germany}

\author{H. Schneider, G. Jakob}
\address{Institut f\"ur Physik, Johannes Gutenberg-Universit\"at Mainz,
Staudinger Weg 7, D-55128 Mainz, Germany}

\author{K. Postava}
\address{Department of Physics, Technical University of Ostrava,
17. listopadu 15, 708 33, Ostrava-Poruba, Czech Republic}

\author{C. Felser}
\address{Institute of Inorganic and Analytical Chemistry, Johannes
Gutenberg-Universit\"at Mainz, Staudingerweg 9, D-55128 Mainz, Germany}

\date{\today}

\begin{abstract}

Co$_2$FeSi(100) films with L2$_1$ structure deposited onto
MgO(100) were studied exploiting both longitudinal (LMOKE) and
quadratic (QMOKE) magneto-optical Kerr effect. The films exhibit a
huge QMOKE signal with a maximum contribution of up to
\unit[30]{mdeg}, which is the largest QMOKE signal in reflection
that has been measured thus far. This large value is a fingerprint
of an exceptionally large spin-orbit coupling of second or higher
order. The Co$_2$FeSi(100) films exhibit a rather large coercivity
of 350 and \unit[70]{Oe} for film thicknesses of 22 and
\unit[98]{nm}, respectively. Despite the fact that the films are
epitaxial, they do not provide an angular dependence of the
anisotropy and the remanence in excess of 1\% and 2\%,
respectively.

\end{abstract}

\maketitle

\section{Introduction}

Since its discovery by John Kerr in 1877 \cite{kerr} the
magneto-optical Kerr effect (MOKE) has evolved into a very
powerful investigational tool for magnetic materials. In most
cases only first order contributions of MOKE are detected, i.e.,
only the contributions linearly proportional to the sample
magnetization. In such a case (neglecting exchange bias systems),
the measured MOKE loops are symmetric. The first order MOKE is
divided into polar MOKE (PMOKE), which is proportional to the
out-of-plane magnetization component, longitudinal MOKE (LMOKE),
and transverse MOKE (TMOKE), with the two latter effects being
proportional to the in-plane magnetization components parallel and
perpendicular to the plane of light incidence, respectively.
\cite{hubert}

However, for some materials, for example Fe/MgO, \cite{pos97},
 Fe/Ag, \cite{cow97} and Py/Ta, \cite{mat99}
or even in diluted magnetic semiconductors as Ga$_{1-x}$Mn$_x$As,
\cite{moo03} asymmetric MOKE loops were observed. Osgood {\it et
al.} \cite{ost98} and Postava {\it et al.} \cite{pos97} first
attributed the asymmetric nature of such loops to a superimposed
quadratic MOKE (QMOKE) contribution, exhibiting an even dependence
on the applied magnetic field. Osgood {\it et al.} originally
stated that the QMOKE signal is related only to the magnetization
term $M_L M_T$, \cite{ost98} and therefore concluded that QMOKE
loops may only appear for systems where the magnetization reversal
occurs by a coherent magnetization rotation, where $M_L$ and $M_T$
are the longitudinal and transverse magnetization components with
respect to the plane of light incidence, respectively. However, it
has been shown later that the QMOKE signal is proportional to two
mixed terms, $M_LM_T$ and $M_L^2-M_T^2$. \cite{pos02} Hence, QMOKE
contributions to magnetization curves may appear also in systems
where magnetization reversal takes place by nucleation and growth
of magnetic domains, as it is the case for the hysteresis loops
presented in this article.

A non-zero Kerr effect requires the presence of both spin-orbit
(SO) coupling and exchange interactions. \cite{hul32,bru96} The
first order MOKE (both PMOKE and LMOKE) originates from a
component of $\vek{M}$ parallel to $\vek{k}$, where $\vek{k}$ is a
wavevector of light \emph{inside} the ferromagnetic materials. In
this configuration, the first order contribution to SO coupling,
$E_\mathrm{SO}=\xi \vek{L}\cdot\vek{S}$, is dominant. \cite{ost98}
On the other hand, the QMOKE originates from a component of
$\vek{M}$ perpendicular to the plane of incidence. In this
situation, the first order of SO coupling is zero, $\xi
\vek{L}\cdot\vek{S}=0$. Therefore, only SO coupling of second or
higher order can give rise to QMOKE. As such higher order
contributions of SO coupling are usually much smaller than those
of first order, the QMOKE is usually much smaller than the
first-order MOKE.

In the present work we investigate the magneto-optical properties
of the half-metallic \CFS\ Heusler compound, which recently
attracted lots of attention as a possible candidate for spintronic
applications. \cite{wur06apl,ino06,kal06,kan06,has06,wur05,tez06,
nic79} This compound exhibits the highest magnetic moment
(\unit[5.97]{$\mu_B$}) per formula unit (at \unit[5]{K},
corresponding to an average value of \unit[1.49]{$\mu_B$} per
atom) and the highest Curie temperature (\unit[1100]{K}) among all
half-metallic ferromagnets investigated so far.\cite{wur06apl} The
magnetic moments per atom derived from a sum rule analysis are
\unit{2.6}$\mu_B$ for Fe and \unit[1.2]{$\mu_B$} for Co at
\unit[300]{K} and \unit[400]{Oe}.\cite{wur05} For those reason, a
thorough characterization of the magnetic properties of this
material is of great interest from a technological point of view.
However, the magneto-optical properties of Co$_2$FeSi have not yet
been studied in detail. Here we perform such an investigation, and
in this context we demonstrate that the material under
investigation exhibits the highest QMOKE that has ever been
measured so far in a magnetic thin film system.

\section{Theory of the longitudinal and quadratic MOKE}

In order to facilitate the understanding of this work, we now
briefly summarize the physical origin and theoretical description
of both first order MOKE and QMOKE.

\begin{figure}
\begin{center}
\includegraphics[width=0.6\textwidth]{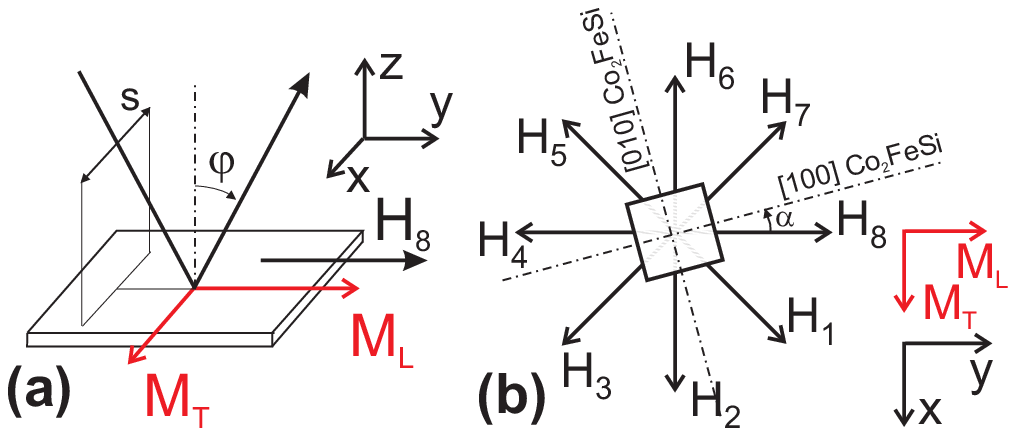}
\end{center}
\caption{%
\label{f:sketch}%
(color online) (a) Sketch of the sample and the incident
s-polarized light. (b) Definition of the 8 directions of the
externally applied magnetic field which are used to determine the
values of the QMOKE signal in saturation. $\alpha$ denotes the
sample orientation, i.e., the angle between the [100] \CFS\
direction and the plane of incidence of the incoming light
($y$-axis). All hysteresis loops presented in this article were
recorded when the positive direction of the magnetic field
corresponds to $\vek{H}_8$ direction, i.e., parallel with the
plane of the incidence of the light.}
\end{figure}

A cartesian coordinate system which will be used throughout this
article is sketched in Fig.~\ref{f:sketch}. The $z$-axis of this
system corresponds to the sample out-of-plane axis, while the
in-plane $y$- and $x$-axes are parallel and perpendicular to the
plane of light incidence, respectively. The sample orientation is
described by an angle $\alpha$, which is the angle between the
$y$-axis and the in-plane [100] direction of the crystal lattice.
In the following, the relative sample magnetization components
along the $y$ and the $x$ axis are called longitudinal $M_L$ and
transversal $M_T$ magnetization, respectively. Note that $M_L$ and
$M_T$ are defined with respect to the plane of light incidence and
not with respect to the orientation of the sample or with respect
to the applied magnetic field.

The optical and magneto-optical properties of a magnetized crystal
are described by the permittivity tensor $\ep_{ij}$, which can be
developed into a series in the
components of the sample magnetization {\bf M}:
\begin{equation}
\label{eq:ep}%
 \ep_{ij}=\ep_{ij}^{(0)}+K_{ijk}M_k+G_{ijkl}M_k
M_l+\cdots,
\end{equation}
where the $M_i$ are the components of {\bf M}. In
Eq.~(\ref{eq:ep}) the Einstein summation convention over the $x$,
$y$, and $z$ coordinates is used. The $\ep_{ij}^{(0)}$, $K_{ijk}$
and $G_{ijkl}$ are constants, forming the dielectric tensor and the linear and
quadratic magneto-optical tensors, respectively.
\cite{vis86} The number of independent components of these tensors
can be reduced using the Onsager relation
\begin{equation}
\label{eq:onsager}%
 \ep_{ij}(\vek{M})=\ep_{ji}(-\vek{M})
\end{equation}
which can also be written in form of general symmetry arguments
\cite{vis86, pos03}
\begin{eqnarray}
\nonumber%
 \ep_{ij}^{(0)}&=\ep_{ji}^{(0)}
\\
\label{eq:ekgsym}%
 K_{ijk}&=-K_{jik}, \quad K_{iik}=0, \quad i\neq j\neq k
\\
\nonumber%
G_{ijkl}&=G_{jikl}=G_{jilk}=G_{ijlk}.
\end{eqnarray}

For cubic crystals (as in the case of \CFS\ studied here), the
number of independent tensor elements can be reduced further,
resulting in only one free (complex) parameter in the constant
term $\ep_{ij}^{(0)}$, another one in the linear term $K_{ijk}$
and three additional parameters in the quadratic term $G_{ijkl}$
(note that the general symmetry relations determined by
Eq.~(\ref{eq:ekgsym}) are still valid) \cite{pos03}
\begin{eqnarray}
\nonumber%
\ep_{ij}^{(0)}&=\ep_d\delta_{ij}\\
\nonumber%
K_{ijk}&=K\\
\label{eq:ekg4}%
G_{iiii}&=G_{11} \\
\nonumber%
G_{iijj}&=G_{12}, \quad i\neq j\\
\nonumber%
G_{1212}&=G_{1313}=G_{2323}=G_{44}~~~,
\end{eqnarray}
with $\delta_{ij}$ being the Kronecker delta-function. Therefore,
the off-diagonal elements of $\ep_{ij}$ can be written as (here we
limit ourselves to in-plane magnetization only):
\begin{eqnarray}
\nonumber%
\ep_{xy}&=\ep_{yx}=2G_{44}M_LM_T\\
\label{eq:epnorot}%
\ep_{xz}&=-\ep_{zx}= K M_L \\
\nonumber%
 \ep_{yz}&=-\ep_{zy}=-K M_T~~~.
\end{eqnarray}
On the other hand, at a sample orientation of $\alpha \neq 0$ the
off-diagonal permittivity tensor elements can be written as
\begin{eqnarray}
\nonumber%
\ep_{xy}&=\ep_{yx}=\left[2G_{44}+\frac{\Delta
G}{2}(1-\cos4\alpha)\right] M_LM_T- \frac{\Delta
G}{4}\sin4\alpha(M_L^2-M_T^2)\\
\label{eq:epgen}%
\ep_{xz}&=-\ep_{zx}= K M_L \\
\nonumber%
 \ep_{yz}&=-\ep_{zy}=- K M_T ~~~,
\end{eqnarray}
by applying a rotational transformation around the $z$ axis to the
permittivity tensor $\ep_{ij}$ (see Refs.~\cite{pos03,vis86a}). In
the above equation, $\Delta G=G_{11}-G_{12}-2G_{44}$ denotes the
so-called magneto-optical anisotropy parameter.

The relation between the complex Kerr amplitude
$\Phi_{s/p}=\theta_{s/p}-i\epsilon_{s/p}$ and the permittivity
tensor elements $\ep_{ij}$ was thoroughly studied in literature.
\cite{hubert,zvezdin} $\theta_{s/p}$ and $\epsilon_{s/p}$ denote
the Kerr rotation and Kerr ellipticity for $s$ and $p$ polarized
incident light, respectively. A general analytical expression of
the Kerr effect is most often too complicated for practical uses.
Therefore two important analytical approximations for ultra-thin
ferromagnetic layers or bulk-like ferromagnets are often used,
assuming either $t_\FM\ll\lambda/4\pi N_\FM$ or
$t_\FM\gg\lambda/4\pi N_\FM$, respectively, with $t_\FM$ and
$N_\FM$ being the thickness and the refractivity index of the FM
layer. In both limiting cases, the analytical solution can be
expressed as \cite{tra92,qiu99,vis95,ham03}
\begin{equation}
\label{eq:kerr0}%
\begin{array}{rl}
\displaystyle%
 \Phi_{s}=-\frac{r_{ps}}{r_{ss}}&=A_s
\left(\ep_{yx}-\frac{\ep_{yz}\ep_{zx}}{\ep_d}\right)+B_s\ep_{zx}
\\[4mm]
\displaystyle%
\Phi_{p}=\frac{r_{sp}}{r_{pp}}&=-A_p
\left(\ep_{xy}-\frac{\ep_{xz}\ep_{zy}}{\ep_d}\right)+B_p\ep_{xz}
\end{array}
\end{equation}
where the weighting optical factor $A_{s/p}$ ($B_{s/p}$) is even
(odd) function of the angle of incidence $\varphi$. If $\ep_{ij}$
from Eq.~(\ref{eq:ekg4}) is substituted to Eq.~(\ref{eq:kerr0}),
we obtain \cite{pos02}
\begin{eqnarray}
\label{eq:kerr1}%
\nonumber \Phi_{s/p}=&\pm A_{s/p}\left[2G_{44}+\frac{\Delta
G}{2}(1-\cos4\alpha)+\frac{K^2}{\ep_d}\right]M_LM_T
\\
& \mp A_{s/p}\frac{\Delta G}{4}\sin4\alpha(M_L^2-M_T^2)\mp
B_{s/p}KM_L,
\end{eqnarray}
where $+$ (-) is related to the Kerr s (p) effect.
Equation~(\ref{eq:kerr1}) is a final expression of the Kerr effect
in case of an in-plane magnetized film with cubic symmetry. It
shows several interesting features. (i) The last term, which is
proportional to $M_L$, describes the ordinary LMOKE. (ii) There
are two separate QMOKE contributions, being proportional to
$M_LM_T$ and $M_L^2-M_T^2$, respectively. (iii) Only the QMOKE
signal strength depends on the crystallographic sample orientation
$\alpha$, characterized by the magneto-optic anisotropy parameter
$\Delta G$. (iv) Even if $G_{ijkl}\equiv0$, there is a small
quadratic contribution proportional to $K^2$. It can be seen from
Eq.~(\ref{eq:kerr0}) that this contribution originates from the
mixed term $\ep_{zx}\ep_{yz}$ ($\ep_{xz}\ep_{zy}$) for $\Phi_s$
($\Phi_p$). This contribution is not due to an intrinsically
quadratic dependence of $\ep_{ij}$ on the sample magnetization,
but arises from a mixing of linear permittivity tensor components.

\section{Experimental details}
\label{s:sample}

\CFS\ films of thicknesses 11, 21, 42 and \unit[98]{nm} were
prepared by RF magnetron sputtering and deposited directly onto
MgO(100). All film thicknesses were measured by means of X-ray
reflectometry, except for the smallest thickness, which was
estimated from sputtering rate and deposition time. All films grow
in the L2$_1$ ordered structure \cite{wur06,schneider06} and were
covered by a \unit[4]{nm} thick Al protective layer. A more
detailed description of the sample preparation process as well as
an investigation of the structural properties of such films can be
found in Ref.\cite{schneider06}. These investigations revealed a
high degree of crystalline order. In particular, no evidence for
significant lattice distortions, such as those reported in
Ref.\cite{has06a} for the growth on GaAs(001) substrates, was
found during these studies.

All Kerr measurements within this article were performed using
s-polarized red laser light of a wavelength of
\unit[$\lambda=670$]{nm}. In each case, a laser spot of
\unit[$\approx$ 300]{\textmu m} diameter was focused onto the
sample surface. All hysteresis loops presented in this article
were recorded when the magnetic field was applied in $\vek{H}_8$
direction (Fig.~\ref{f:sketch}(b)), i.e., parallel with the plane
of the incidence of the light.

\section{Results and discussion}

\subsection{Magnetization curves}

\begin{figure}
\begin{center}
\includegraphics[width=0.5\textwidth]{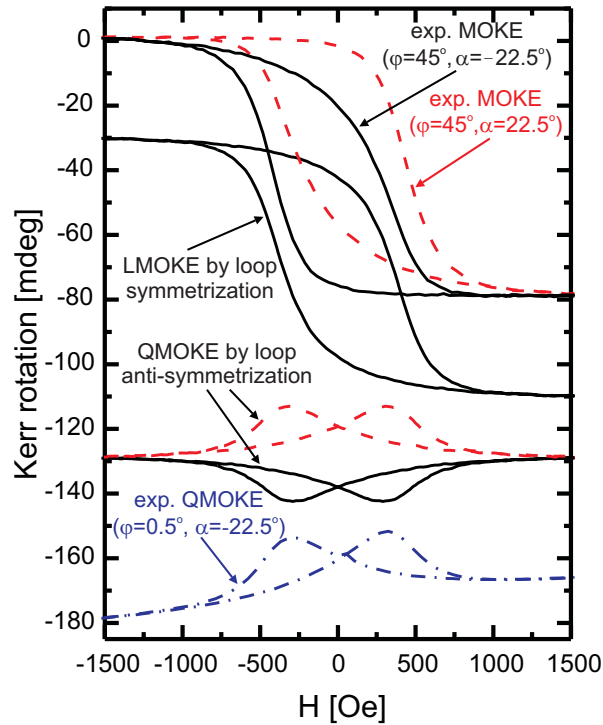}
\end{center}
\caption{%
\label{f:loop}%
(color online) MOKE loops recorded from the \CFS(\unit[21]{nm})
sample. The topmost two loops were directly measured at
$\varphi=45^\circ$ and $\alpha=22.5^\circ$ (red dashed line) or
$\alpha=-22.5^\circ$ (black full line), respectively. These two
loops are symmetrized and anti-symmetrized (see text for details),
providing LMOKE and QMOKE contributions. The bottom loop (blue
dash-doted line) is a QMOKE loop directly measured at
$\varphi=0.5^\circ$ and $\alpha=-22.5^\circ$.}
\end{figure}

Typical Kerr rotation MOKE loops $\theta(H)$ measured on the
\unit[21]{nm} thick sample are presented in Fig.~\ref{f:loop}. The
topmost two loops in Fig.~\ref{f:loop} were measured at an angle
of incidence of $\varphi=45^\circ$ and a sample orientation of
$\alpha=\pm22.5^\circ$. Both loops are asymmetric as they contain
LMOKE (odd in $H$) and QMOKE (even in $H$) contributions.
\cite{pos97,ost98} The loops' asymmetric nature changes sign when
the sample orientation is changed from $\alpha=22.5^\circ$ to
$-22.5^\circ$, corresponding to changes of sign of the QMOKE
contribution. On the other hand, when the sample orientation is
0$^\circ$ or 45$^\circ$ (not shown in Fig.~\ref{f:loop}), the
loops are symmetric, which means that there is no QMOKE
contribution in this case. These results are consistent with the
expected fourfold symmetry of the QMOKE contribution, see
Eq.~(\ref{eq:kerr1}).

Any Kerr rotation loop can be separated into its symmetric
$\theta_{\mathrm{sym}}$ and antisymmetric $\theta_{\mathrm{asym}}$
parts by using the relation:
$\theta_{\mathrm{sym}/\mathrm{asym}}=[\theta_{\mathrm{inc}}(H)\mp\theta_{\mathrm{dec}}(-H)]/2$,
where $\theta_\mathrm{inc/dec}$ denotes the loop branch when $H$
is increasing or decreasing, respectively. A more general way of
the loop symmetrization and antisymmetrization valid also for
systems with exchange bias has been presented by T. Mewes
\textit{et al.} \cite{mew04}

The results of loop symmetrization and antisymmetrization are
visualized in Fig.~\ref{f:loop}. The symmetrized (LMOKE) loops are
identical for $\alpha=\pm22.5^\circ$, so we show only one loop.
Moreover, the antisymmetrized (QMOKE) loops differ only in sign in
this case. Finally, the bottom loop (blue dash-dot line) shows the
QMOKE loop as it is measured directly at a nearly normal angle of
incidence of $\varphi\approx0.5^\circ$ (for $\varphi=0$, the LMOKE
vanishes and hence the measured signal is proportional only to the
QMOKE in the case of in-plane magnetized samples). As expected,
the general shape of this QMOKE loop is identical to that of the
QMOKE loops determined by antisymmetrization of the MOKE loops
measured at $\varphi=45^\circ$. However, the amplitude of the
antisymmetrized QMOKE loop is slightly smaller, which can be
attributed to a reduction of $A_{s}$ when the angle of incidence
increases \cite{vis95}.

\subsection{Amplitude of QMOKE in saturation} \label{s:qmoke}

In the previous section we have shown how to derive a QMOKE loop.
However, it is also interesting to determine the value of the
QMOKE signal in saturation. This parameter is an analogue of the
LMOKE signal in saturation which is experimentally determined (in
the case of Kerr rotation, for example) as
$\theta_{\sat,M_L}=[\theta(H_{\sat})-\theta(-H_\sat)]/2$, where
$H_\sat$ is the saturation field of the sample. The QMOKE signal
in saturation can also be determined, but in a more complicated
way, \cite{pos02} where the Kerr signal is subsequently measured
after application of an external field in eight different
directions $\vek{H}_1$ to $\vek{H}_8$ (Fig.~\ref{f:sketch}(b)),
which is sufficient to saturate the sample each time. In this
case, [see Eq.~(\ref{eq:kerr1})], the LMOKE Kerr rotation signal
in saturation is
$\theta_{\sat,M_L}=[\theta(\vek{H}_8)-\theta(\vek{H}_4)]/2$, while
the corresponding QMOKE signal proportional to $M_LM_T$ and
$M_L^2-M_T^2$ reads
\begin{eqnarray}
\label{eq:8dir1}%
\displaystyle\theta_{\sat,M_LM_T}&
\displaystyle=[\theta(\vek{H}_1)+\theta(\vek{H}_5)-\theta(\vek{H}_3)-\theta(\vek{H}_7)]/4
\\
\label{eq:8dir2}%
\displaystyle\theta_{\sat,M_L^2-M_T^2}&
\displaystyle=[\theta(\vek{H}_8)+\theta(\vek{H}_4)-\theta(\vek{H}_2)-\theta(\vek{H}_6)]/4.
\end{eqnarray}

\begin{figure}
\begin{center}
\includegraphics[width=0.5\textwidth]{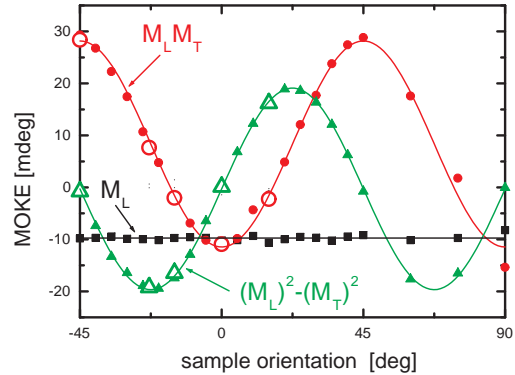}
\end{center}
\caption{%
\label{f:mlmt}%
(color online) (full symbols) Dependence of different MOKE signals
in saturation for the \CFS(\unit[21]{nm}) sample on the sample
orientation $\alpha$ at $\varphi=0.5^\circ$, which are determined
from the 8-directional method \protect\cite{pos02}. (open symbols)
MOKE signals determined from the height of peaks in QMOKE loops.
See text for details.}
\end{figure}

Figure \ref{f:mlmt} displays the angular dependence of the
different MOKE signals in saturation of the \unit[21]{nm} thick
sample at an angle of incidence of $\varphi=0.5^\circ$. Data
determined by the 8-directional method described above are
represented by full symbols. In agreement with
Eq.~(\ref{eq:kerr1}), it is obtained that the LMOKE signal is
independent on $\alpha$ ($\blacksquare$). On the other hand, the
QMOKE signal related to $M_L^2-M_T^2$ is proportional to $\sin
(4\alpha)$ ($\blacktriangle$) whereas the QMOKE signal related to
$M_LM_T$ is proportional to $\cos(4\alpha)+\mathrm{const}$
($\bullet$). The absolute vertical shift in the angular dependence
of $M_LM_T$ is proportional to $G_{44}$ [see Eq.~(\ref{eq:kerr1})]
whereas the amplitudes of both sinusoidal graphs are proportional
to the anisotropy term $\Delta G$.

The experimental data in Fig.~\ref{f:mlmt} exhibit the same
amplitude for both sinusoidal graphs. However,
Eq.~(\ref{eq:kerr1}) shows that there is a factor of 2 between the
strengths of the QMOKE contributions proportional to $M_LM_T$ and
$M_L^2-M_T^2$, respectively. This factor of 2 is effectively
cancelled by the fact that $M_LM_T$ in saturation is given by
$M_LM_T=M^2\cos\alpha\sin\alpha$ and is thus equal to $1/2~M^2$
for $\alpha=45^\circ$.

Finally it can be noted that in the case of the \unit[21]{nm}
thick sample (see Fig.~\ref{f:mlmt}), the QMOKE amplitude is
\unit[20]{mdeg} and the maximal QMOKE signal reaches
\unit[30]{mdeg}. To our knowledge, these values are the highest
QMOKE amplitude and signal in reflection that have ever been
measured.

\subsection{Peak heights in QMOKE loops} \label{s:peak}

\begin{figure}
\begin{center}
\includegraphics[width=0.5\textwidth]{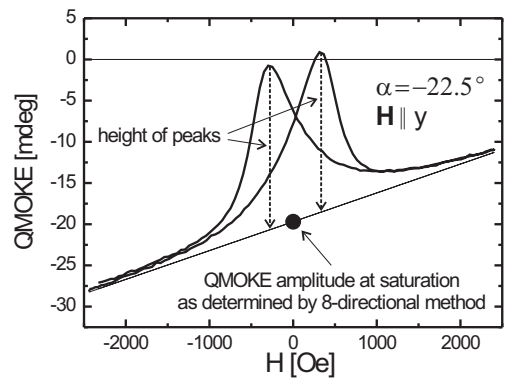}
\end{center}
\caption{\label{f:qmokeloop}%
QMOKE Kerr rotation loop in \CFS(\unit[21]{nm}) measured at
$\varphi=0.5^\circ$, at a sample orientation $\alpha=-22.5^\circ$,
and within positive field ranges applied in $\vek{H}_8\parallel y$
direction. The full circle at $H=0$ shows the QMOKE signal in
saturation as determined by the 8-directional method for this
particular $\alpha$. Dashed arrows illustrate how the height of
the peaks in the QMOKE loop is determined. It can be seen that the
QMOKE reaches its maximum when the sample is in saturation.
Furthermore, the QMOKE is zero when the peaks are reaching their
highest signal.}
\end{figure}

QMOKE loops were measured in the directions of $\vek{H}_1$ to
$\vek{H}_8$ for different sample orientations $\alpha$.
Figure~\ref{f:qmokeloop} shows such an example of a QMOKE loop
determined for the \unit[21]{nm} thick sample at
$\alpha=-22.5^\circ$, when the positive magnetic field was applied
in $\vek{H}_8$ direction, i.e., in $y$-direction. The linear slope
in the QMOKE loops originates from a Faraday effect arising in
optical elements of the experimental setup due to the stray field
of the magnet used for the measurements. The large full circle
($\bullet$) at $H=0$ shows the QMOKE signal in saturation, as
determined in the previous section. The value of this signal
determines the absolute value of Kerr rotation in the QMOKE loop,
which reaches its maxima in saturation. As the magnetic field is
reduced from saturation, the QMOKE signal decreases and finally
reaches zero value at the top of the peaks (i.e., at a field of
$H_c$). Consequently, it holds that $\langle M_L M_T\rangle=0$ and
$\langle M_L^2-M_T^2\rangle=0$ simultaneously during reversal,
i.e., at $H=H_c$, where $\langle\ldots\rangle$ indicates that
these values are averaged over the laser spot area (which has a
diameter of \unit[$\approx$300]{\textmu m}, as noted above). This
fact shows that the reversal process occurs by nucleation and
growth of magnetic domains. In the case of coherent magnetization
reversal, the average values $\langle M_L M_T\rangle$ and $\langle
M_L^2-M_T^2\rangle$ cannot be equal to zero simultaneously. The
presence of magnetic domains during the reversal process is also
confirmed by LMOKE loops measured at $\varphi=45^\circ$, where the
magnetic field was applied in transverse ($x$) direction and no
LMOKE signal was obtained.

In order to check whether all QMOKE loops exhibited zero value at
$H=H_c$, the height of the peaks was determined for loops measured
at different $\alpha$ and at $\vek{H}_1\ldots\vek{H}_8$ directions
of the applied magnetic field, as it is sketched in
Fig.~\ref{f:qmokeloop}. Subsequently, these values were processed
by the 8-directional method. The results obtained in this way are
represented by open symbols in Fig.~\ref{f:mlmt}. It can be seen
that for any $\alpha$ the QMOKE signals determined from peak
heights and related to $M_LM_T$, $M_L^2-M_T^2$ have the same value
as those determined in saturation. Moreover, it can be observed
that for any $\alpha$ the QMOKE loops are always reaching zero
value during reversal at $H=H_c$. Therefore the reversal always
occurs through nucleation and growth of magnetic domains, where
$\langle M_L M_T\rangle=0$ and $\langle M_L^2-M_T^2\rangle=0$.

\subsection{Magnetic anisotropy}

\begin{figure}
\begin{center}
\includegraphics[width=0.8\textwidth]{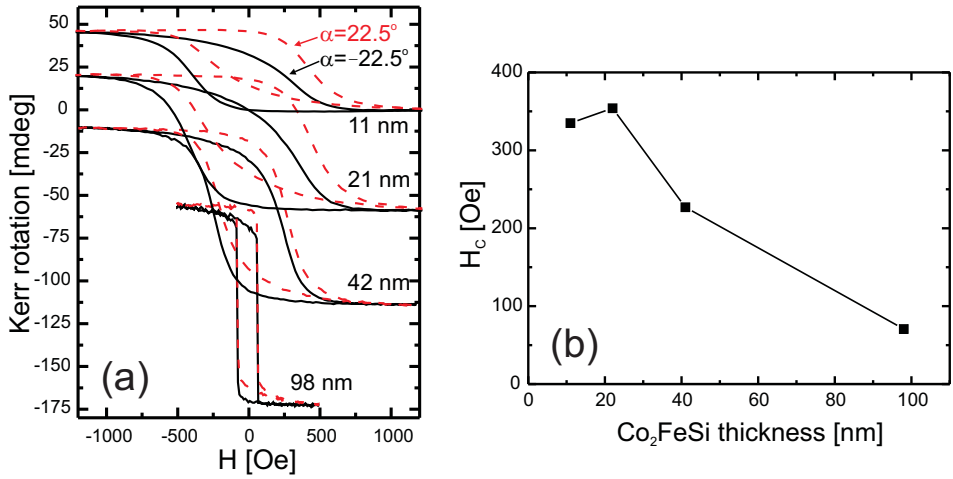}
\end{center}
\caption{%
\label{f:loopthick}%
(color online) (a) MOKE hysteresis loops for different thicknesses
of \CFS\ films, recorded at an angle of incidence
$\varphi=45^\circ$. The sample orientation is equal to
22.5$^\circ$ (dashed red line) or $-22.5^\circ$ (full black line).
(b) Dependence of $H_c$ on Co$_2$FeSi film thickness.}
\end{figure}

Figure \ref{f:loopthick}(a) shows Kerr rotation hysteresis loops
determined for different film thicknesses at two different sample
orientations $\alpha=\pm22.5^\circ$. It can be observed that the
hysteresis loops become more and more squared with increasing
thickness while $H_c$ is decreasing. This can most likely be
attributed to an improvement of the crystalline quality of the
films with increasing thickness, which leads to a higher mobility
of domains walls. The reduction of $H_c$ with increasing thickness
is shown quantitatively in Fig.~\ref{f:loopthick}(b). For this
purpose, the value of $H_c$ was determined from symmetrized LMOKE
loops, as the QMOKE contribution effectively reduces the
coercivity (see Sec.~\ref{s:qmoke}). Figure.~\ref{f:loopthick}(b)
shows that a change of film thickness from \unit[21]{nm} to
\unit[98]{nm} results in a reduction of $H_c$ by almost a factor
of 5, i.e., from \unit[345]{Oe} down to \unit[70]{Oe}.

\begin{figure}
\begin{center}
\includegraphics[width=0.8\textwidth]{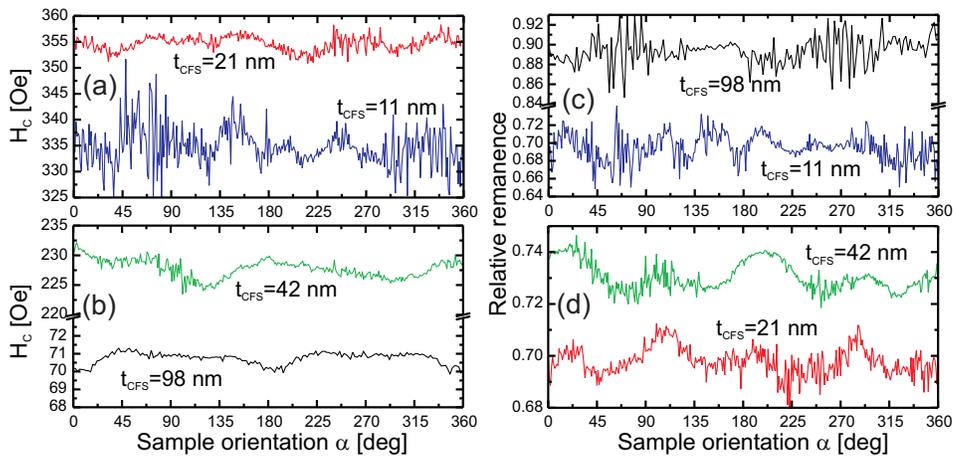}
\end{center}
\caption{%
\label{f:hc}%
(color on-line) Dependence of (a)(b) the coercive field and (c)(d)
the relative remanence on the sample orientation $\alpha$ as
determined from symmetrized LMOKE loops in Co$_2$FeSi films. The
relative remanence is determined as the ratio of the LMOKE signal
at \unit[$H=0$]{Oe} and the LMOKE signal in saturation.}
\end{figure}

Figure~\ref{f:hc}(a) and (b) display the dependence of the
coercive field $H_c$ on the in-plane sample orientation $\alpha$
as determined from LMOKE symmetrized loops for all sample
thicknesses. For film thicknesses of \unit[11]{nm} and
\unit[21]{nm}, the films exhibit a very weak four-fold anisotropy,
modulating $H_c$ by less than 1\%. On the other hand, a film of
\unit[98]{nm} thickness exhibits a weak 2-fold anisotropy, again
modulating $H_c$ by less than 1\%. In the case of a Co$_2$FeSi
film of \unit[42]{nm} thickness a mixture of both a two-fold and a
four-fold anisotropy seem to be present.

A similar situation is shown in Figure~\ref{f:hc}(c) and (d),
which presents the dependence of relative remanence on the sample
orientation $\alpha$. This parameter was determined as a ratio of
the LMOKE signal in zero field normalized by the value of the
LMOKE signal in saturation. In this case, the modulation of the
remanence is about 2\% in magnitude.

The observed very week anisotropy of $H_c$ and remanence is very
surprising, as the samples investigated here are epitaxial and
therefore one should naively expect the presence of hard and easy
axes. However, such hard and easy axes may still be present in our
samples because, as it has been shown in Sec.~\ref{s:peak},
magnetization reversal occurs by domain wall propagation. Hence,
the samples investigated here are unique examples of an epitaxial
system, where the switching field along the hard and easy axis
direction is balanced, resulting in a coercive field that is
nearly independent on the sample orientation $\alpha$.

\subsection{Dependence of MOKE on the film thickness}

\begin{figure}
\begin{center}
\includegraphics[width=0.45\textwidth]{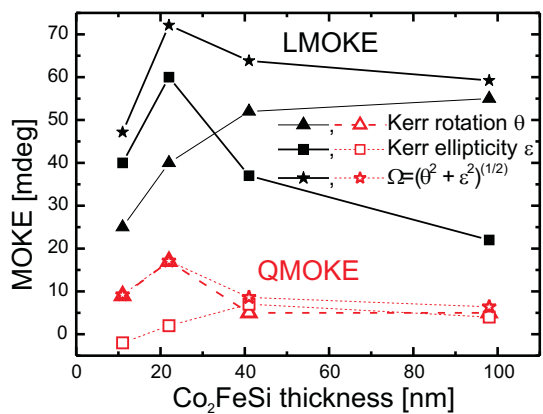}
\end{center}
\caption{%
\label{f:MOKEthick}%
(color online) Dependence of the LMOKE and QMOKE signals at
saturation on the film thickness, measured at $\varphi=45^\circ$.
The QMOKE signal was determined from the height of peaks in QMOKE
loops, which are obtained by antisymmetrization of experimental
MOKE loops.}
\end{figure}

Figure \ref{f:MOKEthick} shows the dependence of different MOKE
signals on the film thickness. We present the Kerr rotation
$\theta$, the Kerr ellipticity $\epsilon$, as well as their
Pythagorean average $\Omega=\sqrt{\theta^2+\epsilon^2}$ for both
the LMOKE and QMOKE signals. Here, the QMOKE signal was determined
from peak heights obtained from measurements recorded at an angle
of incidence $\varphi=45^\circ$ and a sample orientation
$\alpha=22.5^\circ$. It can be observed that the LMOKE signal
($\blacktriangle$) saturates at high thicknesses whereas the QMOKE
signal ($\triangle$) reaches a maximum at an intermediate
thickness and then decreases again. In order to better understand
this behavior, we also measured the Kerr ellipticity in both
cases, which exhibits a similar, but opposite behavior: the LMOKE
Kerr ellipticity ($\blacksquare$) reaches a maximum and then
decreases again whereas the QMOKE Kerr ellipticity ($\square$)
monotonically increases with increasing thickness.

Such a behavior can be attributed to the depth sensitivity of the
magneto-optical Kerr effect.\cite{hubert,tra92,ham02} This is
clearly demonstrated for the Pythagorean average $\Omega$ of both
the LMOKE and QMOKE signals (full and empty stars in
Fig.~\ref{f:MOKEthick}). With increasing film thickness, both
LMOKE and QMOKE Pythagorean averages reach their maxima at a value
in the range of \unit[20-30]{nm} and then saturate with further
increasing thickness.

In particular, the Kerr effect amplitude originating from an
ultrathin sublayer of thickness $\Delta t$ situated at a depth
$t_i$ can be expressed as
\begin{equation}
\label{eq:depth}%
\Phi_i \approx C \Delta t \exp\left[\frac{4\mathrm{i}\pi t_i
N_{z,\subcfs}}{\lambda}\right],
\end{equation}
where
$N_{z,\subcfs}=\sqrt{(N_\subcfs^2-N_\mathrm{air}^2\sin^2\varphi)}$
is the normalized (complex) $k$-vector in $z$-direction,
$N_\mathrm{air}$ is the refractivity index of air and $C$ is a
complex constant.\cite{hubert,tra92,ham02} The resulting Kerr
effect amplitude $\Phi_\mathrm{tot}$ is given by summation over
all contributions originating from different depths,
$\Phi_\mathrm{tot}=\sum_i \Phi_i$. Due to the exponential term in
Eq.~(\ref{eq:depth}), a Kerr signal originating from a deeper
sublayer $t_i$ exhibits a larger damping as well as a larger shift
in phase. The Kerr signals originating from different depths $t_i$
and $t_j$ differ by a phase (Eq.~\ref{eq:depth})
$\Delta\zeta_{i-j}=4\pi\Re(N_{z,\subcfs})(t_i-t_j)/\lambda$. If
the ferromagnetic films are thick and transparent enough, as in
our case, then the Kerr effect amplitudes $\Phi_i$, $\Phi_j$ from
depths $t_i$, $t_j$ may differ by a phase of $\pi$. In such a
case, $\Phi_i$ and $\Phi_j$ cancel each other, leading to a
reduction of the resulting Kerr effect amplitude, in agreement
with the behavior of Pythagorean average
$\Omega=|\Phi_\mathrm{tot}|$ in Fig.~\ref{f:MOKEthick}.

In conclusion, the dependencies of both the Kerr rotation and
ellipticity on the film thickness seem to be determined by
phenomenological optical and magneto-optical properties of the
investigated samples and not by a change of their electronic
structures.

\section{Summary and conclusions}

\CFS\ films in the L2$_1$ structure of thicknesses
11--\unit[98]{nm} and deposited onto MgO(100) were studied by
means of the longitudinal (LMOKE) and quadratic (QMOKE)
magneto-optical Kerr effect. The samples exhibit a huge QMOKE
effect with an amplitude of \unit[20]{mdeg} and a maximum QMOKE
signal reaching \unit[30]{mdeg} at a sample thickness of
\unit[21]{nm}. To our knowledge, these are the highest values of
QMOKE amplitude and signal in reflection that have been measured
so far. For example, for bcc Fe of thickness \unit[50]{nm} and
capped by \unit[1.5]{nm} of Pd, the QMOKE signal is
\unit[5.7]{mdeg}.\cite{pos02} Furthermore, the large QMOKE signal
is a fingerprint of an exceptionally large spin-orbit coupling of
second or higher order in \CFS\ compared to other FM materials
\cite{bru96, ham06cfsion}. It should be noted that the
half-metallicity also contribute significantly to large values of
Kerr effect \cite{ek97}. However, in our case it is also ratio
QMOKE/LMOKE$\approx$0.7 signal which is much larger than in case
of bcc Fe, where this ratio is about 0.1. \cite{pos97}

Moreover, the investigated samples exhibit rather large
coercivities of 350 or \unit[70]{Oe}, corresponding to film
thicknesses of 21 or \unit[98]{nm}, respectively. Although they
are epitaxial, they do not show an angular dependence of the
coercivity as well as the remanence in excess of 1\% or 2\%,
respectively. More detailed investigations of the magnetization
reversal process by means of Kerr microscopy might be required in
order to clarify this issue. The results obtained so far strongly
indicate that such reversal processes take place through
nucleation and growth of magnetic domains. In particular, when the
magnetic field during reversal reaches $H=H_c$, then averaging
over many magnetic domains results in $\langle M_L M_T\rangle=0$
and $\langle M_L^2-M_T^2\rangle=0$ at any sample orientation
$\alpha$. Finally, the thickness dependence of the obtained LMOKE
and QMOKE signals was found to be consistent with a
phenomenological magneto-optical description. Therefore, a
thickness dependence of the electronic structure of the
investigated films could be excluded.

\section{Acknowledgment}

The project was financially supported by the Research Unit 559
\emph{"New materials with high spin polarization"} funded by the
Deutsche Forschungsgemeinschaft, and by the Stiftung
Rheinland-Pfalz f\"ur Innovation. Partial support by NEDO
International Joint Research Program 2004IT093 of the Japanese
government, by European Commission within the EU-RTN ULTRASWITCH
(HPRN-CT-2002-00318) and by Grant Agency of the Czech Republic
(202/06/0531) is gratefully acknowledged. We would like to thank
T.\ Mewes for stimulating discussions.

\newpage
\section{References}

\clearpage

\end{document}